\documentclass[final,5p,times,twocolumn,authoryear]{elsarticle}

\usepackage{amssymb}
\usepackage{lipsum}
\usepackage{amsmath}
\usepackage{newtxtext,newtxmath}
\usepackage[T1]{fontenc}
\usepackage{graphicx}	
\usepackage{ae,aecompl}
\usepackage{ulem}
\usepackage{booktabs}

\journal{Nuclear Physics B}

\begin{document}

\begin{frontmatter}



\title{Exploring the Macroscopic Properties and Nonradial Oscillations of Proto-Neutron Stars: Effects of Temperature, Entropy, and Lepton Fraction}


\author[first]{Sayantan Ghosh}
\ead{sayantanghosh1999@gmail.com}
\author[first]{Shahebaj Shaikh}
\author[first]{Probit J Kalita}
\author[first]{Pinku Routaray}
\author[first]{Bharat Kumar}
\affiliation[first]{organization={Department of Physics and Astronomy, National Institute of Technology Rourkela},
            addressline={}, 
            city={Rourkela},
            postcode={769008}, 
            state={Odisha},
            country={India}}
\ead{kumarbh@nitrkl.ac.in}
\author[sixth]{B.K. Agrawal}
\affiliation[sixth]{organization=
{Saha Institute of Nuclear Physics},
            addressline={ 1/AF Bidhannagar}, 
            city={Kolkata},
            postcode={700064}, 
            state={West Bengal},
            country={India}}
\begin{abstract}
Neutron stars (NSs) have traditionally been viewed as cold, zero-temperature entities. However, recent progress in computational methods and theoretical modelling has opened up the exploration of finite temperature effects, marking a novel research frontier. This study examines Proto-Neutron Stars (PNSs) using the BigApple parameter set to investigate their macroscopic properties. Two approaches are employed: one with constant temperatures (10-50 MeV) and the other fixing entropy per baryon (S) at predefined levels (S = 1 and S = 2). Notably, S remains constant with increasing baryon density due to electron-positron pair formation at finite temperatures. Analysis of PNS mass-radius profiles, considering neutrino trapping and temperature effects, reveals flattened curves and expanded radii with increasing temperature, resulting in slightly higher masses compared to zero temperature. The influence of lepton fraction ($Y_l$) on maximum PNS mass is explored, indicating that higher $Y_l$ values lead to a softer Equation of State (EoS), reducing maximum mass and increasing the canonical radius ($R_{1.4}$). Further investigation of a constant entropy EoS demonstrates that higher entropy is associated with increased maximum PNS masses and flatter mass-radius curves. Central temperature versus maximum mass relationships suggest a correlation between NS mass and temperature. Lastly, we investigate the behaviour of $f$-mode frequencies in PNS. It reveals that the frequency of these modes decreases with increasing entropy and temperature, reflecting complex thermodynamic interactions within the stars.
\end{abstract}




\end{frontmatter}




\section{Introduction}
\label{introduction}

Neutron stars (NSs) cover an extensive density range, roughly several times nuclear density ($\rho_0\sim 10^{14}gm/cc$), throughout their outer crust to their inner core. This density variability presents a significant challenge when formulating a nuclear equation of state (EoS) within the NS, capable of accurately dense matter behaviour under such diverse conditions. Within the realm of cold NSs 
 \citep{JMP2018} 
existing in weak beta equilibrium (a state where weak interactions balance neutron-proton conversion processes), the EoS essentially links pressure and energy density. However, this simplification falls short in addressing the complex and dynamic scenarios that NSs are involved in. In dynamic scenarios like core-collapse supernovae 
\citep{BA2013, CC2014, HTJ2016, M20}, 
proto-neutron stars (PNS) 
\citep{HS2016,AB1986, LV2004, Pons_1999, PRAKASH19971, Kumar2020}, and binary neutron star (BNS) mergers 
\citep{PhysRevD.83.124008,PhysRevD.86.064032,Kaplan_2014,PhysRevC.96.045806,Lalit2019,Beznogov_2020,PhysRevD.102.043006,PhysRevD.103.083012,PhysRevD.104.063016,Koliogiannis_2021},
the equation of state becomes highly temperature-dependent. Following the core bounce during a supernova explosion, a PNS emerges. This entity is extremely hot and lepton-rich. Within the intense heat and density of the PNS, temperatures reach levels where a finite-temperature EoS comes into play \citep{PhysRevC.106.025801, Raithel_2019, PhysRevD.100.066027, PhysRevC.24.1191, SHEN1998435}. This EoS accounts for interactions among its constituents and provides insights into pressure, density, and other thermodynamic quantities of matter within the PNS across varying temperatures. As a result of the extreme conditions, neutrinos, abundantly produced in this environment, are initially trapped \citep{PhysRevD.74.123001, Tripathy2012} within the PNS, interacting with particles like neutrons, protons, and electrons. As the PNS cools and evolves, the finite temperature EoS helps describe the changes in its structure and composition over time. This phase is crucial in shaping the star's subsequent evolution and the surrounding environment's dynamics. A PNS may transform into a NS or further contract into a black hole (BH) depending on the temperature and other variables. The temperature plays a crucial role in determining the fate of the PNS. If the temperature is high enough, the PNS can retain sufficient thermal energy to prevent further gravitational collapse, leading to the formation of a stable NS. However, if the temperature is too low, the PNS may continue to collapse and eventually produce a BH.

When two NSs merge, such as observed in the case of GW170817, which is considered the source of gravitational wave phenomena known as kilonovae, it results in the creation of an intensely energetic and hot environment. The collision and subsequent processes generate temperatures that can sustain a finite-temperature EoS temporarily, crucial for understanding the behaviour of the resulting remnant. This remnant could either form a hypermassive NS or promptly collapse into a black hole, with a finite-temperature EoS being crucial for accurate description under conditions of extreme heat and pressure. The EoSs at finite temperature play a significant role in modelling and comprehending the structure, composition, and evolution of NSs during critical events like supernova explosions and NS mergers \citep{LV2004, BAIOTTI2019103714, 10.3389/fspas.2020.609460, chabanov2023impact, PhysRevLett.124.171103, Abbott_prl2017}.
 Understanding the dynamic evolution of NSs \citep{glendenning2012compact, doi:10.1126/science.1090720, Blaschke2018} necessitates grasping the implications of finite temperature. Processes like accretion from companion stars \citep{Wang_2016}, nuclear reactions \citep{Lau_2018}, and thermal radiation emission \citep{Potekhin_2014} are fundamental astrophysical phenomena involving NSs. Including limited temperature in theoretical models enhances accuracy in depicting these processes and aids in understanding known NS events.
The LS model proposed by Lattimer and Swesty in 1991 \citep{LATTIMER1991331}, relying on the finite-temperature compressible liquid drop theory with a Skyrme nuclear force, and the relativistic mean field (RMF) theory with a Thomas–Fermi approximation developed by Shen et al. in 2011 \citep{Shen_2011} are prominent examples of such models.

Unlike previous studies \citep{PhysRevC.24.1191, PhysRevC.100.054335, galaxies10040079, PhysRevD.107.103054} that assumed a uniform temperature distribution within NSs, our work addresses the complex interplay between temperature, entropy, density, and pressure in extreme environments. NSs are characterized by significant density, pressure, and degeneracy effects that overshadow the influence of temperature \citep{LATTIMER1991331}. Our investigation centres on the EoS of NSs at finite temperatures, advocating for a constant entropy approach. This approach provides a more accurate representation of the physical conditions, particularly in scenarios like supernova explosions and NS mergers, where temperature fluctuations play a substantial role,
affecting their EoS \citep{Das_2021, Kochankovski_2022}. Macroscopic properties, such as the mass-radius relationship, dimensionless tidal deformability ($\Lambda$) \citep{PhysRevD.81.123016}, fundamental ($f$-mode) frequency \citep{Zhao_2022, PhysRevC.103.035810, Pradhan_2022}, and cooling behavior \citep{Kumar2020,10.1093/mnras/stab3126} of NSs are significantly influenced.

This paper is organized as; 
In Sub-Section (\ref{DM}), we describe the temperature-dependent EoS corresponding to the Lagrangian density for the Relativistic mean field (RMF) model and the BigApple's parameter values. In Sub-Sections (\ref{TOV}), (\ref{fmode}), and (\ref{TD}), we discuss the theoretical background of TOV equations (Hydrostatic equilibrium), $f$-mode oscillations of NSs and Tidal Deformability, respectively. We present our results and discussion part in Section (\ref{RD}), which includes the effect of temperature and entropy on NS properties. And finally in Section (\ref{summary}) we conclude our work.

Throughout this paper, we adopt mostly positive signatures (-, +, +, +) and utilize a geometrized unit system (G=c=1).
\section{Formalism}

\label{formalism}
\subsection{Temperature dependent Equation of state }
\label{DM}
In our study, we take into account matter composed of baryons and leptons at certain temperatures $T$, baryon density $\rho_{B}$, and lepton fraction $Y_{l}$. The exchange of different mesons serves as a model for the interaction of baryons in the relativistic field theory~\citep{WALECKA1974491, BOGUTA1977413, Serot_1997, PhysRevC.90.044305}. 
We have incorporated and adapted this formalism from earlier research \citep{Kochankovski_2022}.
The Lagrangian for a temperature-dependent RMF model can be written as:
\begin{eqnarray}
{\cal L} &=& \sum_{b} \bar{\Psi}_{b} \Bigl\{\gamma_{\mu}(i\partial^{\mu}-q_{b} A^{\mu}-g_{\omega b}\omega^{\mu} -g_{\phi b}\phi^{\mu}
 \nonumber \\
&-& g_{\rho b}\vec{I}_b \, \vec{\rho \, }^{\mu} )  
- (m_{b} - g_{\sigma b}\sigma)
\Bigl\} \Psi_{b}  \nonumber \\
&+& \sum_{l} \bar{\Psi}_{l}\left(i\gamma_{\mu}\partial^{\mu}-q_{l}\gamma_{\mu}A^{\mu}
-m_{l}\right )\Psi_{l} \nonumber \\
&+& \frac{1}{2}\partial_{\mu}\sigma \partial^{\mu}\sigma
-\frac{1}{2}m^{2}_{\sigma}\sigma^{2} - \frac{\kappa}{3!} (g_{\sigma N}\sigma)^3
 \nonumber \\
&-& \frac{\lambda}{4!} (g_{\sigma N}\sigma)^4 \nonumber \\
&-& \frac{1}{4}\Omega^{\mu \nu} \Omega_{\mu \nu} +\frac{1}{2}m^{2}_{\omega}\omega_{\mu}\omega^{\mu}  + \frac{\zeta}{4!}   (g_{\omega N}\omega_{\mu} \omega^{\mu})^4 \nonumber \\
&-&\frac{1}{4}  \vec{R}^{\mu \nu}\vec{R}_{\mu \nu}+\frac{1}{2}m^{2}_{\rho}\vec{\rho}_{\mu}\vec{\rho \, }^{\mu} 
 \nonumber \\
&+& \Lambda_{\omega} g_{\rho N}^2 \vec{\rho}_{\mu}\vec{\rho \,}^{\mu} g_{\omega N}^2 \omega_{\mu} \omega^{\mu} \nonumber \\
&-&\frac{1}{4}  P^{\mu \nu}P_{\mu \nu}+\frac{1}{2}m^{2}_{\phi}\phi_{\mu}\phi^{\mu} -\frac{1}{4} F^{\mu \nu}F_{\mu \nu} 
\label{eq:L}
\end{eqnarray}

where the first term includes baryonic contributions $( b = n, p )$ , the second term includes leptonic contributions $( l = e, \mu )$   and  rest part including contributions  from the $\sigma , \omega , \rho , \phi$ mesons with corresponding neutrinos.
Also, $\Psi_b$ and $\Psi_l$ are the baryon and lepton Dirac fields, respectively. The mesonic and electromagnetic strength tensors can be given as $\Omega_{\mu \nu} = \partial_{\mu} \omega_{\nu} -\partial_{\nu} \omega_{\mu} $, $\vec{R}_{\mu \nu} = \partial_{\mu} \vec{\rho_{\nu}} - \partial_{\nu} \vec{\rho_{\mu}} $, $P_{\mu \nu} = \partial_{\mu} \phi_{\nu} -\partial_{\nu} \phi_{\mu} $ and $F_{\mu \nu} = \partial_{\mu} A_{\nu} -\partial_{\nu} A_{\mu}$. 
Finally, the $\vec I_b$ stands for the isospin operator, the $\gamma_{\mu}$ denotes the Dirac matrices.
The electromagnetic couplings are indicated by $q$, the masses of the baryons, mesons, and leptons by $m$, and the strong interaction coupling of a meson to a specific baryon is marked by $g$ (with $N$ signifying nucleon).
It is important to note that all octet baryon interactions can be mediated by the $\sigma$, $\omega$, and $\rho$ mesons, but the  $\phi$ mesons can only mediate interactions between octet baryons with non-zero strangeness. The above Lagrangian coupling constants approximately encode the intricate nuclear many-body dynamics.  The ground-state features of finite nuclei are crucially dependent on the $g_{\sigma N}$ and $g_{\omega N}$ couplings of the isoscalar $\sigma$ and $\omega$ mesons to the nucleon, which respectively define the energy per particle and density of the nuclear matter saturation point. The nuclear symmetry energy is dependent on the $g_{\rho N}$ coupling of the isovector $\rho$ meson to the nucleon. Symmetry energy is a measure of the amount of work required to convert all the protons in a nucleus into neutrons. Heavy neutron-rich nuclei and NSs are affected by the $g_{\rho N}$ interaction~\citep{PhysRevC.89.044001}.\\
According to Boguta \& Bodmer \citep{BOGUTA1977413}, the Lagrangian density (\ref{eq:L}) takes into account the self-interactions of the meson fields. Boguta \& Bodmer~\citep{BOGUTA1977413} presented the $\sigma$-meson self-interactions, which include the $\kappa$ and $\lambda$ couplings. The EoS softens up at moderate densities due to these couplings, and the realistic compressibility of nuclear matter is obtained, agreeing with experimental observations of nuclear giant resonances and heavy ion collisions. Moreover, Bodmer~\citep{BODMER1991703} introduced the quartic self-coupling $\zeta$ of the vector $\omega$ meson. The attractive nonlinear interaction that softens the EoS at high densities, directly affecting the maximum mass and structure of NSs, is caused by the nonnegative $\zeta$ coupling, which prevents abnormal solutions of the vector field equation of motion~\citep{BODMER1991703, MULLER1996508}. Finally, the neutron radii of heavy nuclei and the radii of NSs are affected by a mixed interaction between the $\omega$ and $\rho$ mesons, exhibiting the coupling $\Lambda_{\omega}$~\citep{Tolos_2017}.
First, using the Lagrangian density given by Eq. (\ref{eq:L}), one can derive the Dirac equations for the various baryons and leptons, which are then used to calculate the thermodynamic properties and composition of matter at a finite temperature under $beta$-equilibrium conditions:
\begin{eqnarray}
&&(i\gamma_{\mu}\,\partial^{\mu}-q_{b}\,\gamma_{0}\,A^{0}-m^{*}_{b} \nonumber \\
&&-g_{\omega b}\, \gamma_{0} \, \omega^{0} -g_{\phi b} \,\gamma_{0}\, \phi^{0} 
-g_{\rho b}\, I_{3 b}\, \gamma_{0} \,\rho_3^{0}) \Psi_{b}=0 , \nonumber \\ 
&&\left(i\gamma_{\mu}\,\partial^{\mu}-q_{l}\,\gamma_{0} \,A^{0}-m_{l} \right) \Psi_{l}=0 , \label{MFlep}
\label{eq:Dirac_Eq}
\end{eqnarray}
with the effective mass of the baryons provided by 
\begin{equation}
m_b^{*} = m_b - g_{\sigma b} \sigma, 
\label{eq:meff}
\end{equation}
and $I_{3b}$ being the third component of the isospin of a particular baryon, with the convention that for protons $I_{3p} = +1/2$. Due to the assumption of rotational invariance and charge conservation, only the time-like component of the vector fields and the third component of isospin have been written in Eq.(\ref{eq:L}).
Serot and Walecka describe in their publication~\citep{Serot_1997} how to derive the equations of motion for mesons using the Euler-Lagrange equations. When these equations are solved, a system of interrelated, non-linear, and strongly-coupled equations is produced. A complex quantization of meson and baryon fields is needed to solve these equations properly. The ability of mesons to interact with baryons in the nuclear medium is significant. Replacing the expectation values of the operators in the meson field allows us to solve these equations. To further understand how these expectation values affect the motion of baryons, we may model them as classical fields. This technique is commonly known as the relativistic mean-field theory.\\
The meson mean fields in the uniform matter can be represented as
$\bar \sigma= \langle \sigma \rangle$, $\bar\omega=\langle\omega^0\rangle$, $\bar\rho=\langle\rho_3^0\rangle$,
and $\bar \phi=\langle\phi^0\rangle$,
In the uniform medium, the mean-field approximation yields the following mesonic equations of motion:
\begin{eqnarray}
&&m_\sigma^2 \, \bar \sigma + \frac{\kappa}{2} g_{\sigma N}^3 \bar \sigma^2 + \frac{\lambda}{3!}  g_{\sigma N}^4 \bar \sigma^3 = \sum_{b} g_{\sigma b} \rho_b^s , \nonumber \\ 
&& m_\omega^2 \, \bar \omega + \frac{\zeta}{3!}  g_{\omega N}^4 \bar \omega^3 + 2 \Lambda_{\omega} g_{\rho N\,}^2  g_{\omega N}^2  \bar \rho^2 \bar \omega = \sum_{b} g_{\omega b} \rho_b , \nonumber \\ 
&& m_\rho^2 \,  \bar \rho + 2 \Lambda_{\omega} g_{\rho N}^2  g_{\omega N}^2  \bar \omega^2 \bar \rho= \sum_{b} g_{\rho b} I_{3 b} \rho_b , \nonumber \\ 
&& m_\phi^2 \bar \phi \, = \sum_{b} g_{\phi b} \rho_b ~, \label{eqphi}
\end{eqnarray}
where the densities at finite temperature, in scalar $\rho_b^s$ and vector $\rho_b$ form, are provided by
\begin{equation}
\begin{aligned}
\label{eq:bary_dens}
\rho_b= & <\bar{\Psi}_b \gamma^0 \Psi_b> \\
= & \frac{\gamma_b}{2 \pi^2}\int_0^{\infty}\!\! dk \, k^2 \, (f_{b}(k,T) - f_{\bar{b}}(k,T))  , \hspace{8cm} 
\end{aligned} 
\end{equation}
\begin{equation}
\begin{aligned}
\rho_b^s= & <\bar{\Psi}_b\Psi_b>   \\
= & \frac{\gamma_b}{2 \pi^2}\int_0^{\infty} \!\! dk \, k^2 \, \frac{m^*_b}{\sqrt{k^2+m_b^{*2}}}\left(f_{b}(k,T)+f_{\bar{b}}(k,T)\right), \hspace{8cm}
\end{aligned} 
\end{equation}
with $\gamma_b=2$ representing the degeneracy of the baryon spin degree of freedom~\citep{PhysRevC.97.045806,PhysRevC.100.054314,PhysRevC.61.054904} and
\begin{equation}
f_{b/\bar{b}}(k,T) =\left[1+\text{exp}\left(\frac{\sqrt{k^2+m_b^{*2}} \mp \mu_b^{*}}{T}\right)\right]^{-1}
\label{eq:distribution}
\end{equation}
being the Fermi-Dirac distribution for the baryon ($b$) and antibaryon ($\bar b$) with effective mass $m_b^*$ and the accompanying effective chemical potential provided by
\begin{equation}
\mu_b^{*} = \mu_b - g_{b\omega}\bar \omega  - g_{b\rho} I_{3b} \bar \rho- g_{b\phi}\bar \phi.
\label{eq:mueff}
\end{equation}
NS cores are composed of globally neutral, $\beta$-equilibrium matter~\citep{SHEN1998435, Stone}.
Therefore, the conditions in the core of a NS are related to the chemical potentials and the number densities of the various particles as 
\begin{eqnarray}
&&\mu_n - \mu_p = \mu_e - \mu_{\nu_e},\nonumber \\
&&0=\sum_{b,l} q_i \, \rho_i \ ,  \nonumber \\
&& \rho=\sum_{b} \rho_i \ ,
\label{beta-eq}
\end{eqnarray}
where  $q_i$ stands for the charge of particle $i$ also $\mu_n , \mu_p , \mu_e$ and $\mu_{\nu_e}$ are chemical potentials of neutron, proton, electron and neutrino respectively.
Finally, conservation holds for the densities of baryons and leptons.
\begin{equation}
 Y_l \cdot \rho_B = \rho_{l}+ \rho_{\nu_l}  \nonumber \\
\end{equation}
\begin{equation}
\rho_{l (\nu_l)}=\frac{\gamma_{l(\nu_l)}}{2 \pi^2}\int_0^{\infty} dk \, k^2 \, {(f_{l (\nu_l)}(k,T) - f_{\bar{l} (\bar{\nu}_l)}(k,T) )}
\end{equation}
$\rho_B$ represents the overall density of baryons, $Y_l$ represents the fraction of leptons with a particular flavour, and $\rho_{l(\nu_l)}$ represents the density of leptons (or leptonic neutrinos)~\citep{2017IJMPD..2650077Z, Pons_1999, PRAKASH19971,2018arXiv180101350C, PhysRevLett.66.2701}. The densities of these particles are described using the Fermi-Dirac distribution, denoted by $f_{l(\nu_l)} (k,T)$ for leptons and neutrinos and their respective antiparticles, represented by $f_{\bar l(\bar \nu_l)} (k,T)$. It is important to remember that the lepton number is no longer conserved in the extreme neutrino-free case, when $\mu_{\nu_l}=\mu_{\bar\nu_l}=0$. \\
Self-consistent solutions are sought for the Dirac equations (\ref{MFlep}) involving the baryons and leptons and the field equations (\ref{eqphi}) involving the mesonic fields $\sigma$, $\omega$, $\rho$, and $\phi$ for a given total baryon density $n$. 

The energy density and pressure of the NS matter can be calculated from the chemical potential and species density at a given $n$~\citep{PhysRevC.89.044001, PhysRevD.104.063016}.
From the stress-energy momentum tensor, the other thermodynamic variables can be easily calculated~\citep{Fetter, Kochankovski_2022}.
\begin{equation}
T_{\mu \nu} =  \frac{\partial {\cal L}}{\partial (\partial_{\mu}\Phi_{\alpha})}\partial_{\nu}\Phi_{\alpha} - \eta_{\mu \nu}{\cal L}. 
\end{equation}
The energy density $\epsilon$ can be calculated as:
\begin{eqnarray}
\label{eq:energy}
\epsilon &=& \langle T_{00}\rangle  \nonumber \\
&&=\frac{1}{2\pi^2}\sum_{b} \gamma_b \int_0^{\infty} dk k^2\sqrt{k^2 + m_b^{*2}}\,(f_{b}(k,T) \nonumber \\
&& + f_{\bar{b}}(k,T)) \nonumber \\
&& +  \frac{1}{2\pi^2} \sum_{l} \gamma_l \int_0^{\infty} dk k^2\sqrt{k^2 + m_l^{2}}\, (f_{l}(k,T) \nonumber \\
&& + f_{\bar{l}}(k,T)) \nonumber \\
&& + \frac{1}{2}(m_{\omega}^2 \bar \omega^2+m_{\rho}^2\bar \rho^2+m_{\phi}^2\bar \phi^2+ m_{\sigma}^2\bar\sigma^2+m_{\sigma^{*}}^2\bar \sigma^{{*}^2}) \nonumber \\
&& +\frac{\kappa}{3!}(g_{\sigma N}\bar \sigma)^3 + 
\frac{\lambda}{4!}(g_{\sigma N}\bar \sigma)^4  \nonumber \\
&& + \frac{\zeta}{8}(g_{\omega N}\bar \omega)^4 + 3\Lambda_{\omega}(g_{\rho N}g_{\omega N}\bar \rho \bar \omega)^2
\end{eqnarray}
Using the thermodynamic relation, we can finally determine the pressure.
\begin{equation}
P=\sum_{i}\mu_{i}\rho_{i}-\epsilon 
\label{press}
\end{equation}
The entropy density~\citep{Prakash_1997,10.1143/PTP.100.1013} is calculated by
\begin{align}
s= \sum_i & \frac{\gamma}{2\pi^2} \int_0^{\infty} dk\,k^2\,[-f_{b}(k,T)\ln f_{b}(k,T) \nonumber  \\ 
- & \left(1-f_{b}(k,T)\right)\ln \left(1-f_{b}(k,T)\right) \nonumber \\ 
- & f_{\bar{b}}(k,T)\ln f_{\bar{b}}(k,T) \nonumber \\
- & \left(1-f_{\bar{b}}(k,T)\right) \ln \left(1-f_{\bar{b}}(k,T)\right)].
\label{Eq:entropy}
\end{align}
Here we have defined the Entropy per baryon (S) as, S=$s$/n, where n is the baryon number density.

In our work, we used BigApple parameterization obtained by Fattoyev et al.\citep{PhysRevC.102.065805} which reproduces the $2.6M_{\odot}$ mass of NS considering nucleons and extended it to finite temperature. The coupling constants and corresponding empirical nuclear properties at saturation for the BigApple parameter set are given in Table \ref{tab:BigApple}.
\begin{table}
  \centering
  \setlength{\tabcolsep}{17 pt}
  \caption{Parameters of the model BigApple~\citep{PhysRevC.102.065805}. The mass of the nucleon is equal to $m_N = 939$ MeV.}
    \begin{tabular}{ccc}
    \toprule \hline
    Parameter & Value & Unit \\
    \midrule 
    $m_{\sigma}$ & 492.73 & MeV \\
    $m_{\omega}$ & 782.5 & MeV \\
    $m_{\rho}$ & 763.0 & MeV \\
    $g_{\sigma N}^2$  & 93.506 & - \\
    $g_{\omega N}^2$ & 151.68 & - \\
    $g_{\rho N}^2$ & 200.556 & - \\
    $\kappa$ & 5.20326 & MeV \\
    $\lambda$ & - 0.021739 & - \\
    $\zeta$ & 0.00070 & - \\
    $\Lambda_{\omega}$ & 0.047471 & - \\
    \noalign{\smallskip}\hline
    \hline
    \end{tabular}
  \label{tab:BigApple}
\end{table}
\subsection{Hydro-static Equilibrium Structure}
\label{TOV}
The Tolman-Oppenheimer-Volkoff (TOV) equations, which describe the hydrostatic equilibrium of NSs, are implied by Einstein's field equations in Schwarzschild-like coordinates~\citep{PhysRev.55.374, PhysRev.55.364}.
For temperature-dependent EoS, TOV equations allow us to quickly determine the mass and radius of the static isotropic  PNS by applying boundary conditions. The TOV equations are:
\begin{align}
\frac{dP}{dr} &= - \frac{\left( P+\cal E \right) \left( m + {4\pi r^3P}\right)}{r^2{\left(1-\frac{2m}{r}\right)}} \, ,
\nonumber \\
\frac{dm}{dr} &= {4\pi r^2 \cal E} \, ,
\label{eq:TOV}
\end{align}
Integrating these equations from $r = 0$ where $m(r = 0) = 0$ and $P (r = 0) = P_c $ (the central pressure) upto the stellar surface $r = R$ where $m(r = R) = M$ and $P (r = R)=0 $. We solve the TOV equations for temperature-dependent BigApple EoS to get the NS mass-radius profile. 
\subsection{$f$-mode Oscillations of NSs}
\label{fmode}
GWs of various frequencies, including the fundamental $f-$mode, are emitted when NSs experience oscillations due to external or internal perturbations~\citep{PhysRevD.66.104002}. The nonradial oscillations of spherically symmetric NSs can be solved by using the Cowling approximation~\citep{10.1093/mnras/101.8.367}, which assumes that the spacetime is frozen, allowing the metric perturbation to be ignored. The oscillations' eigenvalues are real, signifying their absence of damping. This unique characteristic allows us to investigate these oscillations as purely harmonic modes, as previously established by  \citep{1990ApJ...348..198L}. Consequently, these modes have been extensively studied in the context of neutron stars (NSs) and play a pivotal role in unveiling their internal structure and dynamic behaviour.
\\
The Lagrangian displacement vector of the fluid is given by \\
\begin{equation}
\xi^{i}=\frac{1}{r^2}\Big(e^{-\lambda (r)}W (r),-V
(r)\partial_{\theta},
     -\frac{V(r)}{ \sin^{2}{\theta}}\  \partial _{\phi}\Big)
e^{i\omega t}Y_{lm}(\theta,\phi)
\end{equation}
 where $Y_{lm}(\theta,\phi)$ represents the spherical harmonics. The Lagrangian, defined by the functions $V(r)$ and $W(r)$ and governed by the frequency $\omega$, is determined through the solution of the following system of ordinary differential equations\citep{PhysRevD.83.024014},
\begin{eqnarray}
\frac{d W(r)}{dr}&=&\frac{d {\cal E}}{dP}\left[\omega^2r^2e^{\lambda
(r)-2\Phi (r)}V(r)
+\frac{d \Phi(r)}{dr} W (r)\right] \nonumber \\
&&
-l(l+1)e^{\lambda (r)}V (r) \nonumber \\
\frac{d V(r)}{dr} &=& 2\frac{d\Phi (r)}{dr} V
(r)-\frac{1}{r^2}e^{\lambda (r)}W (r).
\label{eqn:cowling}
\end{eqnarray}
Where, $\Phi(r)$ and $\lambda(r)$ represent metric functions. 
With the fixed background metric
\begin{equation}
\label{eq:metric}
ds^2=-e^{2 \Phi}d t^2+e^{2 \lambda} d r^2+r^2 (d \theta^2+\sin ^2 \theta d \phi^2),
\end{equation}
In the close vicinity of the origin, the solution to Equation \ref{eqn:cowling} exhibits the following behaviour:
\begin{equation}
     W (r)=Br^{l+1}, \ V (r)=-\frac{B}{l} r^l,
\label{eq:bc1}
\end{equation}
where $B$ is an arbitrary constant. To ensure that the perturbation pressure becomes zero at the outer boundary of the star's surface, we need to apply the following additional boundary condition,
\begin{equation}
     \omega^2 e^{\lambda (R)-2\Phi (R)}V (R)+\frac{1}{R^2}\frac{d\Phi
(r)}{dr}\Big|_{r=R}W (R)=0.
\label{eq:bc2}
\end{equation}
Utilizing the boundary conditions outlined in Eq. \ref{eq:bc1} and Eq. \ref{eq:bc2}, we can successfully solve Eq. \ref{eqn:cowling} and determine the eigenfrequencies of the neutron star.
\subsection{Tidal Deformability}
\label{TD}
When a NS is subjected to the gravitational influence of an external field $\left(\epsilon_{ij}\right)$ generated by a companion object, it develops a quadrupole moment $\left(Q_{ij}\right)$. This quadrupole moment is directly proportional to the tidal field and can be represented as follows \citep{Hinderer_2008, Hinderer_2009}:
\begin{equation}
Q_{i j}=-\alpha \epsilon_{i j}.
\end{equation}
The parameter $\alpha$ characterizes the tidal deformability of the star and can be expressed in terms of the tidal Love number ($k_{2}$) as $\alpha = \frac{2}{3} k_{2} R^{5}$, where $R$ denotes the radius of the star. \\

To calculate the tidal Love number of a neutron star (NS), a linear perturbation is introduced to the background metric, following the approach described in previous works such as \citep{Hinderer_2008, Damour_Tidal}. This involves investigating the $tt$-component of the metric perturbation denoted as $H$ in the perturbed Einstein equation. The behaviour of this perturbation is governed by the following second-order differential equation: 
\begin{equation}
 H^{\prime \prime}+H^{\prime} \mathrm{U} +H \mathrm{V}  =0 .
\end{equation}
where;

\begin{align*}
    \mathrm{U} &= \left[\frac{2}{r}+e^{2\lambda}\left(\frac{2 m(r)}{r^{2}}+4 \pi r(P-\mathcal{E})\right)\right] \\
    \mathrm{V} &= \left[4 \pi e^{2\lambda}\left(4 \mathcal{E}+8 P+\frac{\mathcal{E}+P}{d P / d \mathcal{E}}\left(1+c_s^{2}\right)\right)-\frac{6 e^{2\lambda}}{r^{2}}-4\Phi^{\prime^{2}}\right]
\end{align*}
where $\Phi^{\prime}$ is,
\begin{align}
\frac{d \Phi}{d r} &= -\frac{1}{P+ \cal{E}} \frac{d P}{d r}
\end{align}
The determination of the tidal Love number $(k_2)$ involves considering the regularity of $H$ within the NS core and ensuring the continuity of $H(r)$ and its derivative at the surface~\citep{Hinderer_2008, Damour_Tidal}.
\begin{equation}
\begin{aligned}
k_2= & \frac{8}{5} C^5(1-2 C)^2\left[2\left(y_2-1\right) C-y_2+2\right] \\
& \times\left\{\left[\left(4y_2+4\right) C^4+\left(6 y_2-4\right) C^3-\left(22 y_2-26\right) C^2\right.\right. \\
& \left.+3\left(5 y_2-8\right) C-3\left(y_2-2\right)\right]2 C  +3(1-2 C)^2 \\
& \left.\times\left[2\left(y_2-1\right) C-y_2+2\right] \log (1-2 C)\right\}^{-1}
\end{aligned} 
\end{equation}
using the definition of $y_2$ as where $C$ is the compactness of the anisotropic NS provided by $C= \frac{M}{R}$
\[
y_2=\left.\frac{r H^{\prime}}{H}\right|_{at \ r=R} .
\]
Tidal deformability ($\Lambda$), which has no dimensions, is a valuable characteristic that may be derived from GW data~\citep{Choi2019}. The dimensionless tidal deformability of the NS ($\Lambda$) may be computed after we know the values of $k_2$ and $C$, using the formula:
\begin{equation}
    \Lambda= \alpha / M^{5} = \frac{2}{3} k_{2} C^{-5}
\end{equation}

\section{Results and Discussions}

\label{RD}
\begin{figure}
    \centering
    \includegraphics[width=3.45in]{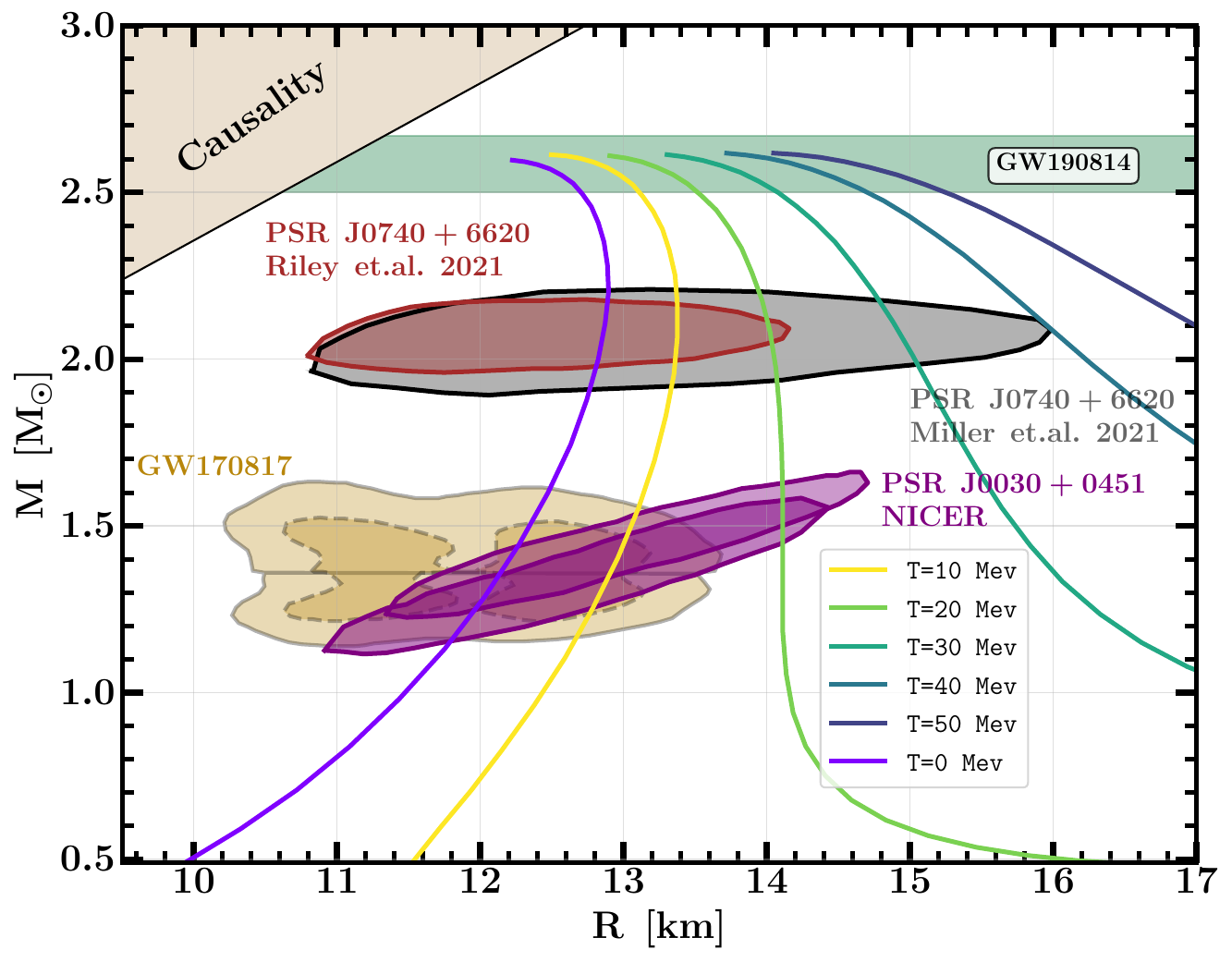}
    \caption{Mass-Radius relations of the PNS for a fixed temperature EoS with $Y_l$=0.0, for BigApple parameter set. The figure shows the recent observational limit on the maximum mass of the NS from GW190814 and the colored shaded regions represent the findings from accurately measured NS mass and radius, such as  PSR J0740+6620 NICER(2021)\citep{Fonseca_2021, Riley_2021} and PSR J0030+0451 NICER(2019)\citep{Riley_2019, Chen_2020} respectively. The outer and inner regions of the golden butterfly structured plot indicate the 90\% (solid) and 50\% (dashed) confidence intervals based on the LIGO-Virgo analysis for Binary Neutron Star (BNS) components of the GW170817 event\citep{Abbott_prl2017, PhysRevLett.120.172702, PhysRevC.107.055804}.}
    \label{MR_T}
\end{figure}

\begin{figure}
    \centering
    \includegraphics[width=3.5in]{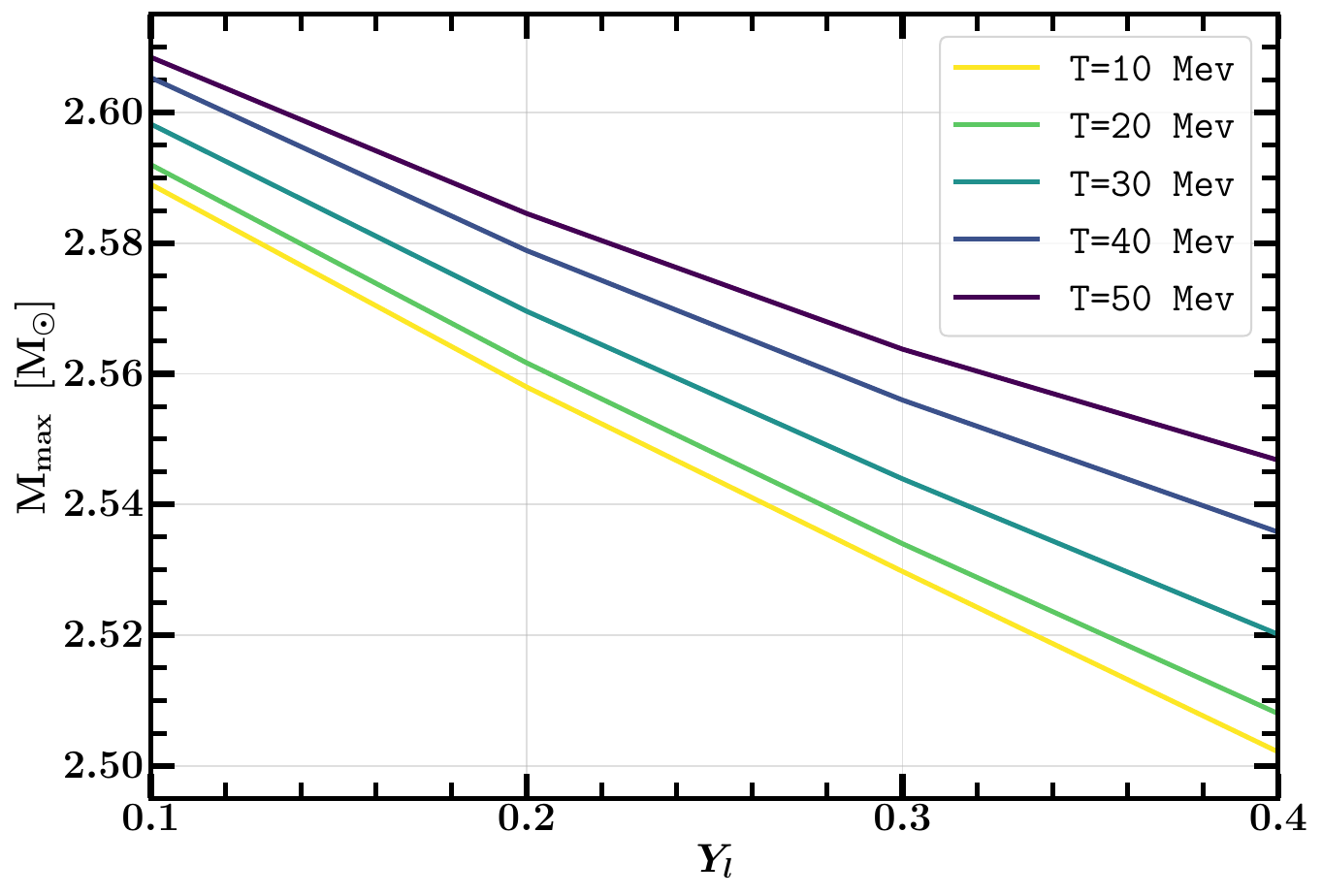}
    \caption{Variation of M$_\mathrm{{max}}$ with lepton fraction $Y_{l}$ for BigApple EoS at different constant temperatures in range T=10-50 MeV indicated by respective colour lines.}
    \label{M_Yl}
\end{figure}

\begin{figure}
    \centering
    \includegraphics[height=7cm,width=8.68cm]{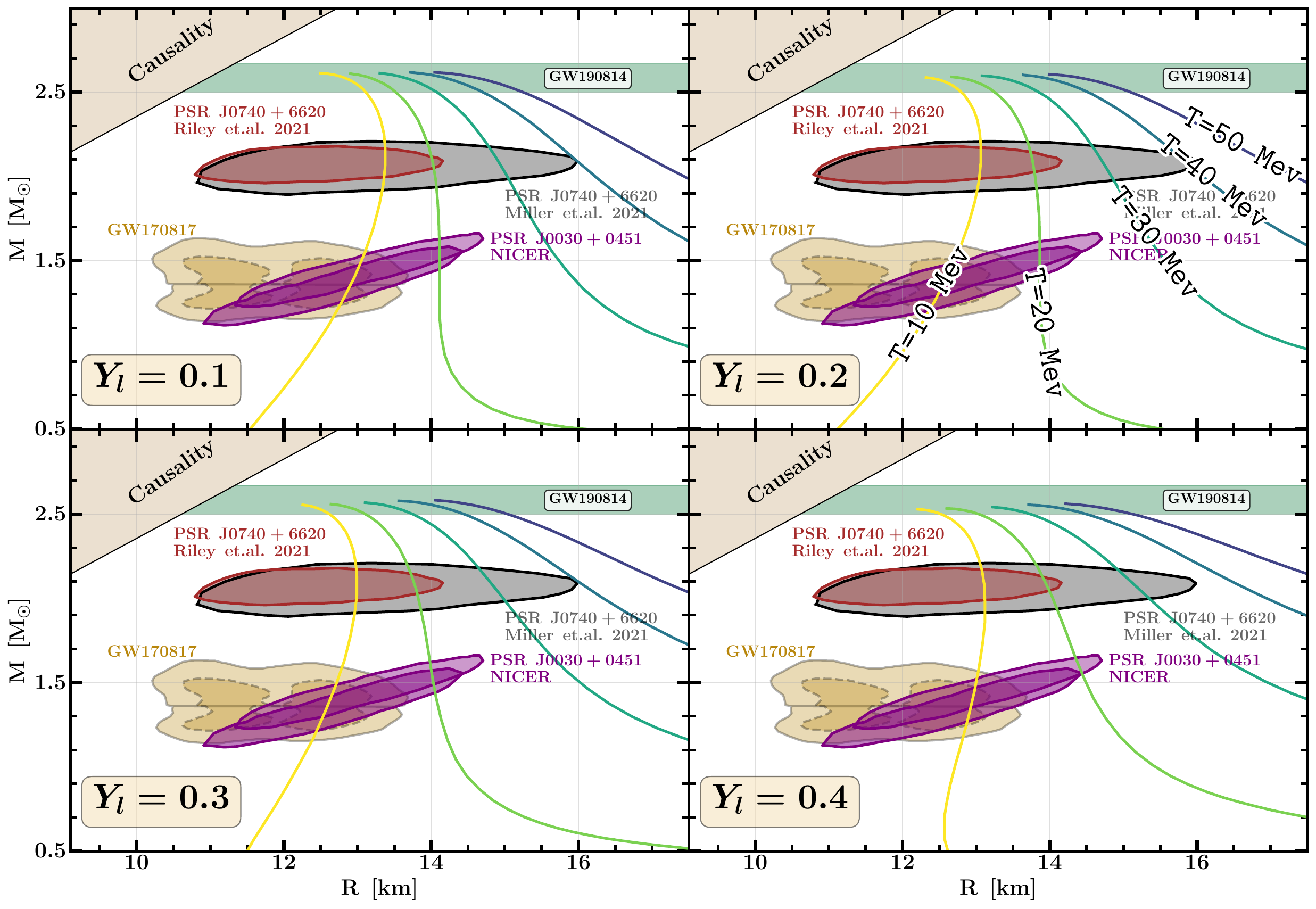}
    \caption{ This figure is the same as previous Fig.\ref{MR_T}, but here with each fixed temperature EoS, we are varying the lepton fraction. The four subplots correspond to values of lepton fraction $Y_l$= 0.1-0.4.}
    \label{MR_T_all}
\end{figure}

We employ two distinct approaches to investigate the macroscopic properties of PNS using the BigApple parameter set. The first approach involves maintaining a constant temperature (T= 10, 20, 30, 40 \& 50 MeV) throughout the star, while the second approach involves holding the entropy per baryon (S) of the NS matter at one of two predetermined values (S = 1 and S = 2). In our calculations for the constant entropy EoS, we adopt the value of $Y_l$ = 0.1, 0.2, 0.3, 0.4. It is important to note that the assumption of a constant temperature is applicable mainly to the lower end of the temperature spectrum. If the temperature within the NS interior rises above the critical Fermi temperature, the star may eventually become unstable~\citep{Kumar2020}.
\\
To study the behaviour of the entropy per baryon (S) in a constant temperature EoS, we employed Eq.\ref{Eq:entropy} to calculate it over a range of temperatures (T = 10-50 MeV) and plotted the results with respect to baryon density.
\begin{figure}
    \centering
    \includegraphics[height=7cm,width=8.68cm]{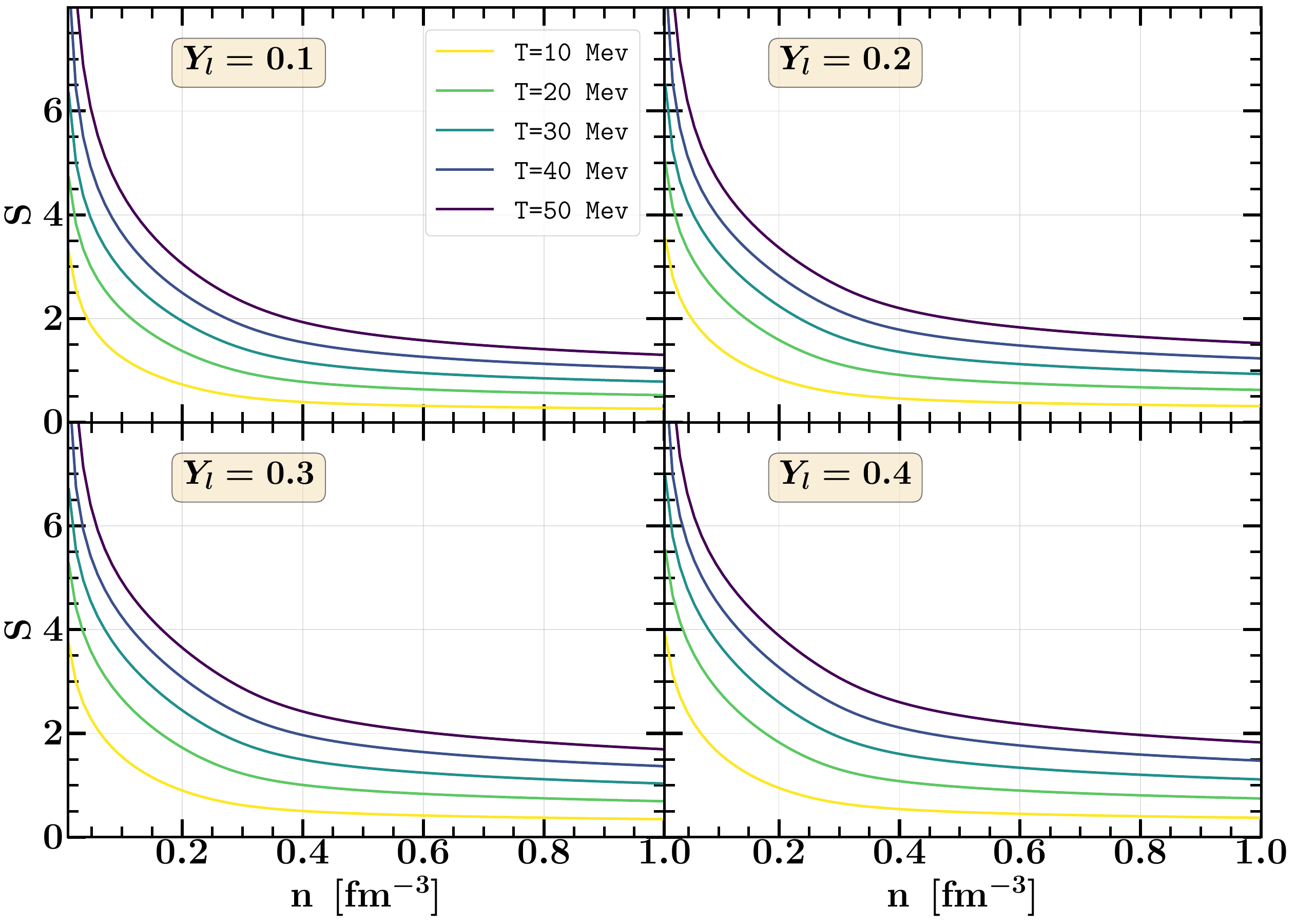}
    \caption{Variation of Entropy per baryon (S) with baryon number density (n) for constant temperature EoS at different values of temperatures from 10-50 MeV represented by respective colour lines. The four subplots correspond to values of lepton fraction $Y_l$= 0.1-0.4.}
    \label{Fig:S_nb}
\end{figure}
The resulting plot, between the entropy per baryon (S) and baryon density (n) with varying the lepton fraction ( $Y_l$ = 0.1 - 0.4 ), as shown in Fig.\ref{Fig:S_nb}, reveals intriguing behaviour: the entropy per baryon (S) remains constant with the increase of baryon density, except for small densities.
This unique trend can be attributed to the presence of electron-positron pairs, which form spontaneously due to thermal fluctuation at finite temperatures. When electron-positron pairs form, they introduce additional ways (i.e. microstates) in which particles can be arranged. These additional microstates increase the total available microstates in the system. So according to Boltzmann's formula, the system's entropy even at very low baryon densities also exists. This finding is consistent with the results reported in~\citep{Dexheimer_2008}, reinforcing the significance of thermal effects on the entropy of dense nuclear matter. 
In Fig \ref{MR_T}, we present the mass-radius profile of the PNS obtained using a fixed temperature EoS with neutrino trapping, along with the most recent observational constraint on the maximum mass of NS. The BigApple parameter set follows to the constraints imposed by the GW190814 observation~\citep{PhysRevC.102.065805} across the entire assumed temperature range. Our analysis reveals that the inclusion of neutrino trapping causes the M-R curve  flatten at the top, significantly affecting the radius of the PNS. Furthermore, the consideration of temperature leads to increased pressure at a given baryon density, resulting in a notable increase in the radius of the PNS. Consequently, the PNS exhibits a slightly larger mass compared to a NS at zero temperature as shown in Fig \ref{MR_T}, primarily due to the EoS being stiffer in the former case.
Next, we examine the influence of the lepton fraction $Y_l$ on the maximum mass of a PNS, as depicted in Fig \ref{M_Yl}. The lepton fraction indicates the abundance of leptons, such as electrons and neutrinos, within the stellar matter. Leptons introduce additional pressure-support mechanisms in a PNS, contributing significantly to the pressure that opposes gravitational collapse. As shown in the figure, we observe an intriguing phenomenon: with an increasing lepton fraction, while maintaining a constant temperature, the maximum mass of the PNS decreases. This behavior is a consequence of the enhanced presence of leptons, which directly impacts the EoS governing the PNS's structure. The increasing lepton fraction results in a softer EoS, leading to reduced pressure at a given baryon density, thereby decreasing the maximum mass and increasing the canonical radius of the PNS.
To gain a more comprehensive understanding of this effect, we further investigate the impact of the lepton fraction $Y_l$ in Fig \ref{MR_T_all}. Within the range of $0.1$ to $0.4$, while holding the temperature constant, we explore the lepton fraction's role in shaping NSs. As the lepton fraction increases, the composition of the NS changes, with a higher proportion of leptons in its matter. This change in lepton content significantly influences the EoS governing the NS's structure. A higher lepton fraction leads to a softer EoS, resulting in lower pressure at a given baryon density compared to cases with lower lepton fractions. Consequently, the NS's maximum sustainable mass decreases as the additional pressure support provided by the leptons becomes less effective in countering gravity at higher lepton fractions. On the other hand, as the lepton fraction increases, the canonical radius $R_{1.4}$, representing the radius at a fixed mass, also increases. As a result, the mass-radius profile of the NS becomes flatter, indicating a more extended structure for NSs with varying masses. The flatter mass-radius profile observed at higher lepton fractions results from the interplay between the softening of the EoS due to the presence of leptons and the additional pressure support they offer. The softer EoS makes the interior of the NS less compressible, allowing it to achieve larger radii at the same mass.
 
\begin{figure}
    \centering
    \includegraphics[width=3.5in]{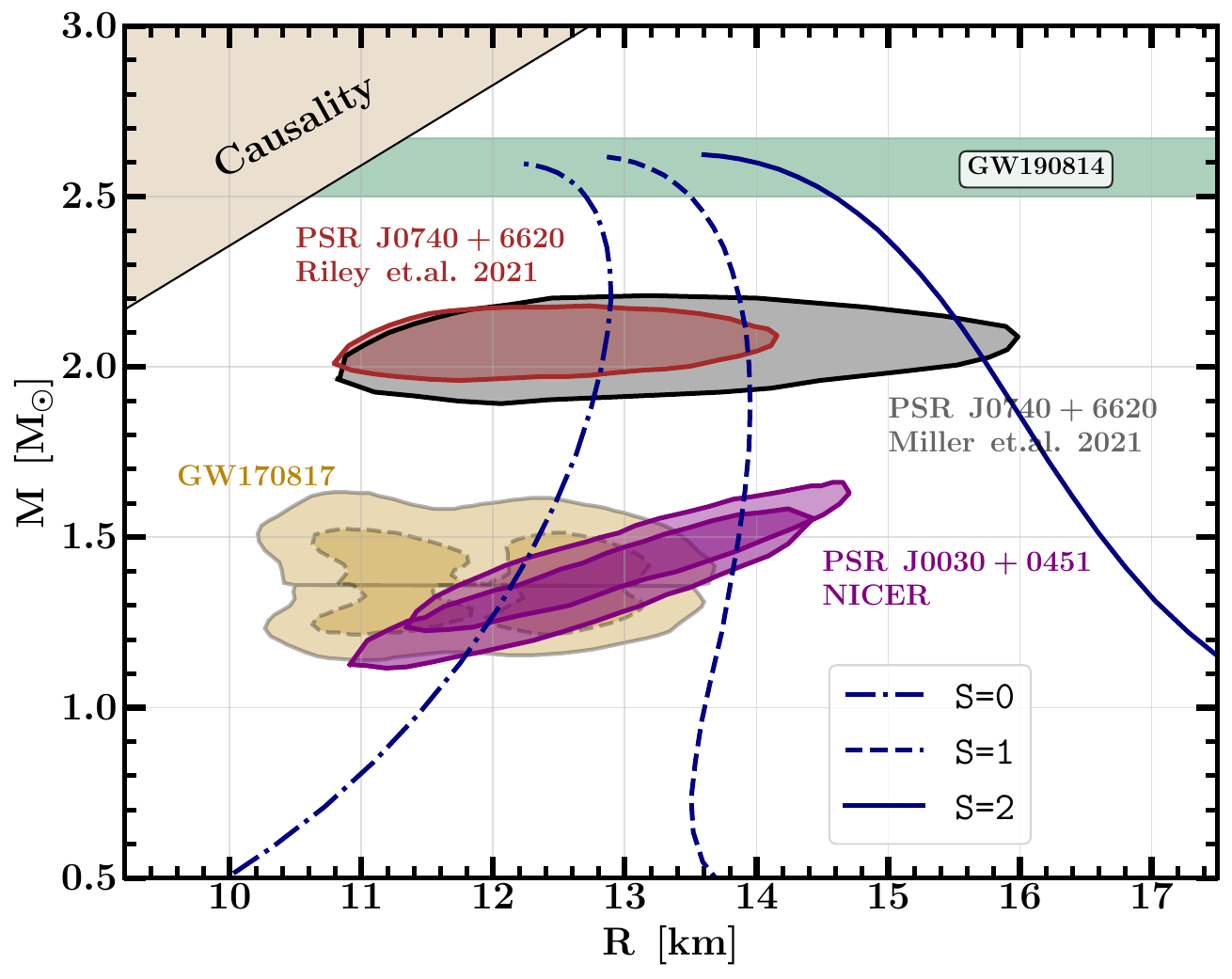}
    \caption{ This figure is the same as previous Fig.\ref{MR_T}, but with constant entropy (S=0, 1, 2) EoS for $Y_l$=0.0.
    }
    \label{MR_s}
\end{figure}

\begin{figure}
    \centering
    \includegraphics[width=3.5in]{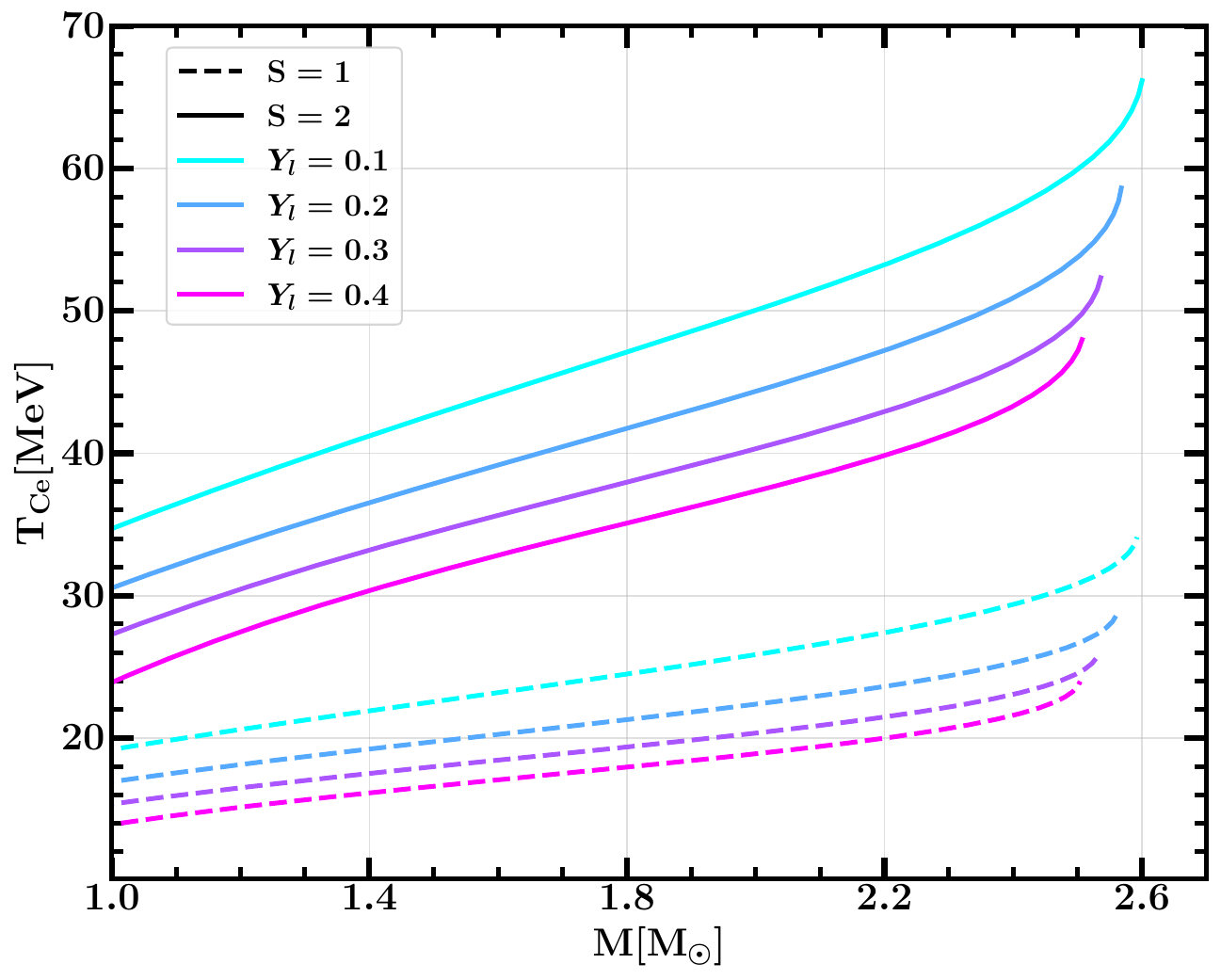}
    \caption{Central Temperature vs Mass [$M_{\odot}$] at a constant entropy S=1(dotted lines) and S=2(solid lines) for different  $Y_{l}$ values i.e. $Y_{l}$=0.1-0.4 represented by respective coloured lines.}
    \label{T_M}
\end{figure}

  To explore NSs matter, it is imperative to take into account the temperature gradient and its influence on their structural properties, while assuming a constant entropy at finite temperatures~\citep{Kumar2020}. By employing the constant entropy EoS, we performed an analysis of the mass-radius profile of a PNS, as depicted in Fig. \ref{MR_s}. The results in the figure showcase an intriguing trend: as the entropy increases, the maximum masses of the PNS rise, and the mass-radius curve becomes notably flatter. This observation reveals a direct correlation between entropy and the stellar structure~\citep{Khadhikar2021}. As the entropy rises, the pressure support within the NS becomes more dominant compared to other forces acting on the PNS. This additional pressure support allows the PNS to exhibit a more extended and less compact structure, ultimately leading to a larger radius for a given mass. The flattening effect on the mass-radius profile arises due to the increased resistance of the PNS to compression, resulting in a less steep mass-radius curve when compared to cases with lower entropy.
\\
Fig. \ref{T_M} depicted the central temperature versus the maximum mass of PNS. From the figure, it is noticed that NS mass increases as the temperature rises, suggesting that more massive stars have hotter cores and play a crucial role in stellar evolution, influencing various processes such as nuclear reactions and energy generation within the star.  The analysis of the mass-radius profile of PNSs in Fig. \ref{MR_s} is also suggests that as the mass of a PNS increases, its radius also increases. In other words, more massive PNSs tend to have larger radii. Additionally, the results in Table \ref{table:MR_s} support this observation, indicating that more massive PNSs have correspondingly larger radii for a constant entropy EoS with central temperature. The results imply that the newborn star's evolutionary process tends toward shrinking in size. This conclusion is based on the fact that massive PNSs, which are hotter and have larger radii, lose mass during their evolutionary phases and cool down. As the PNS cools down, it contracts, leading to a reduction in its size or radius over time.

\begin{figure}
    \centering
    \includegraphics[width=3.5in]{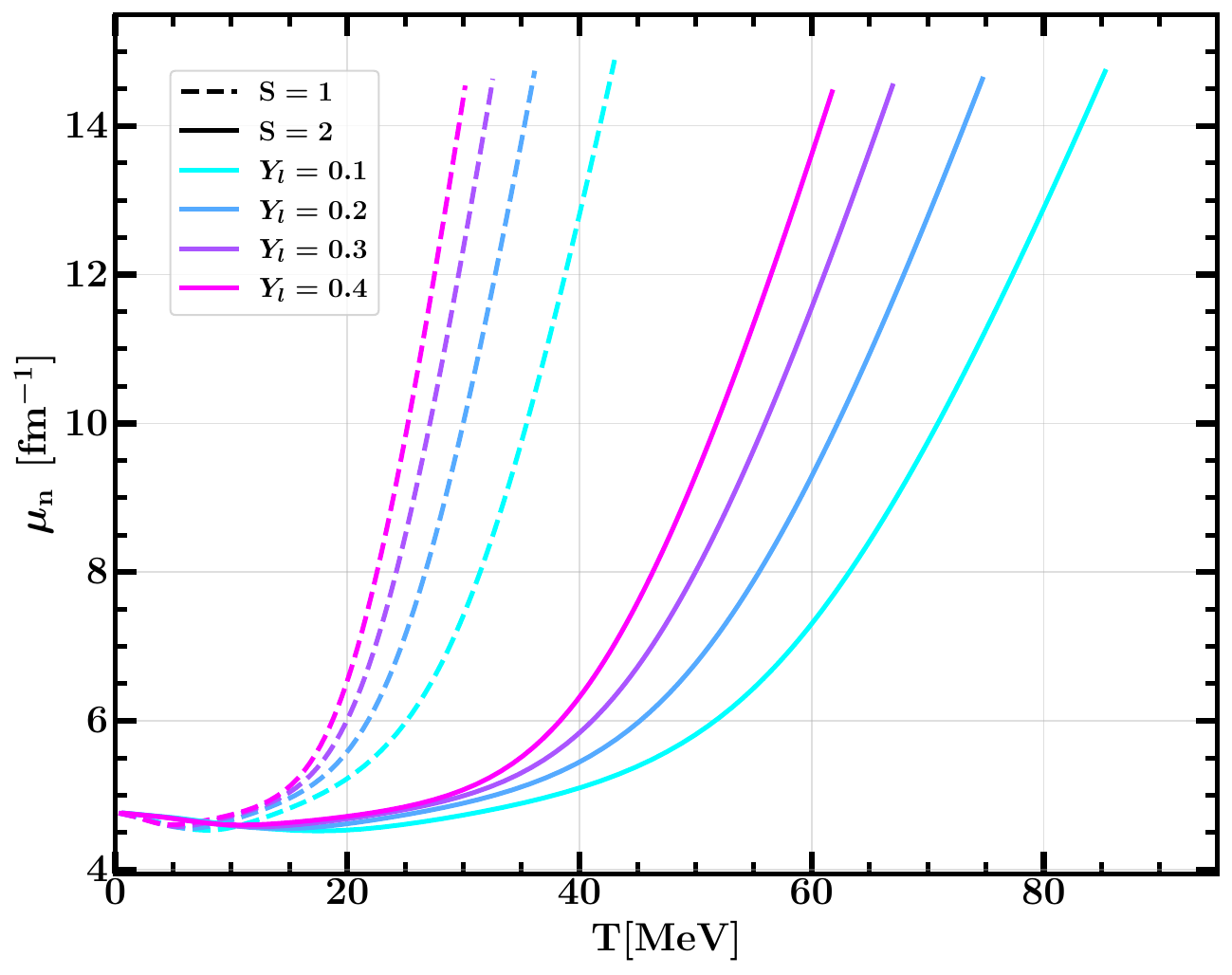}
    \caption{Variation of chemical potential of neutron ($\mu_n$) with temperature at a constant entropy S=1(dotted lines) and S=2(solid lines) for different  $Y_{l}$ values i.e. $Y_{l}$=0.1-0.4 represented by respective coloured lines.}
    \label{T_mu_p}
\end{figure}

\begin{figure}
    \centering
    \includegraphics[width=3.5in]{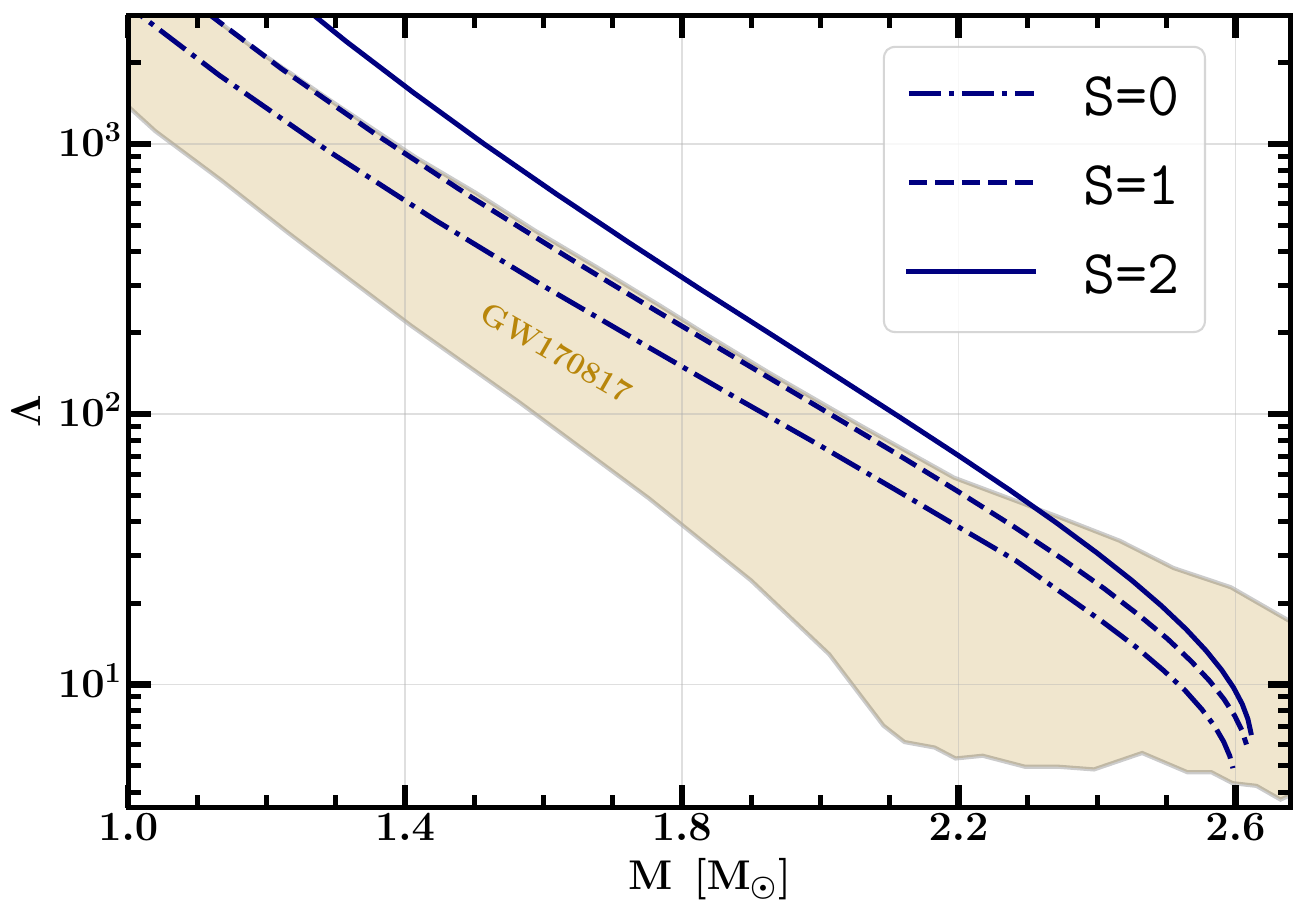}
    \caption{Dimensionless tidal deformability ($\Lambda$) vs Mass [M$_{\odot}$]  for constant entropy values S=0,1 and 2 ($Y_l$=0.0) shown by respective lines. The light brown shaded region is the observational constraints from GW170817 event\citep{Abbott_prl2017}. }
    \label{TD_s012}
\end{figure}

\begin{figure}
    \centering
    \includegraphics[width=3.5in]{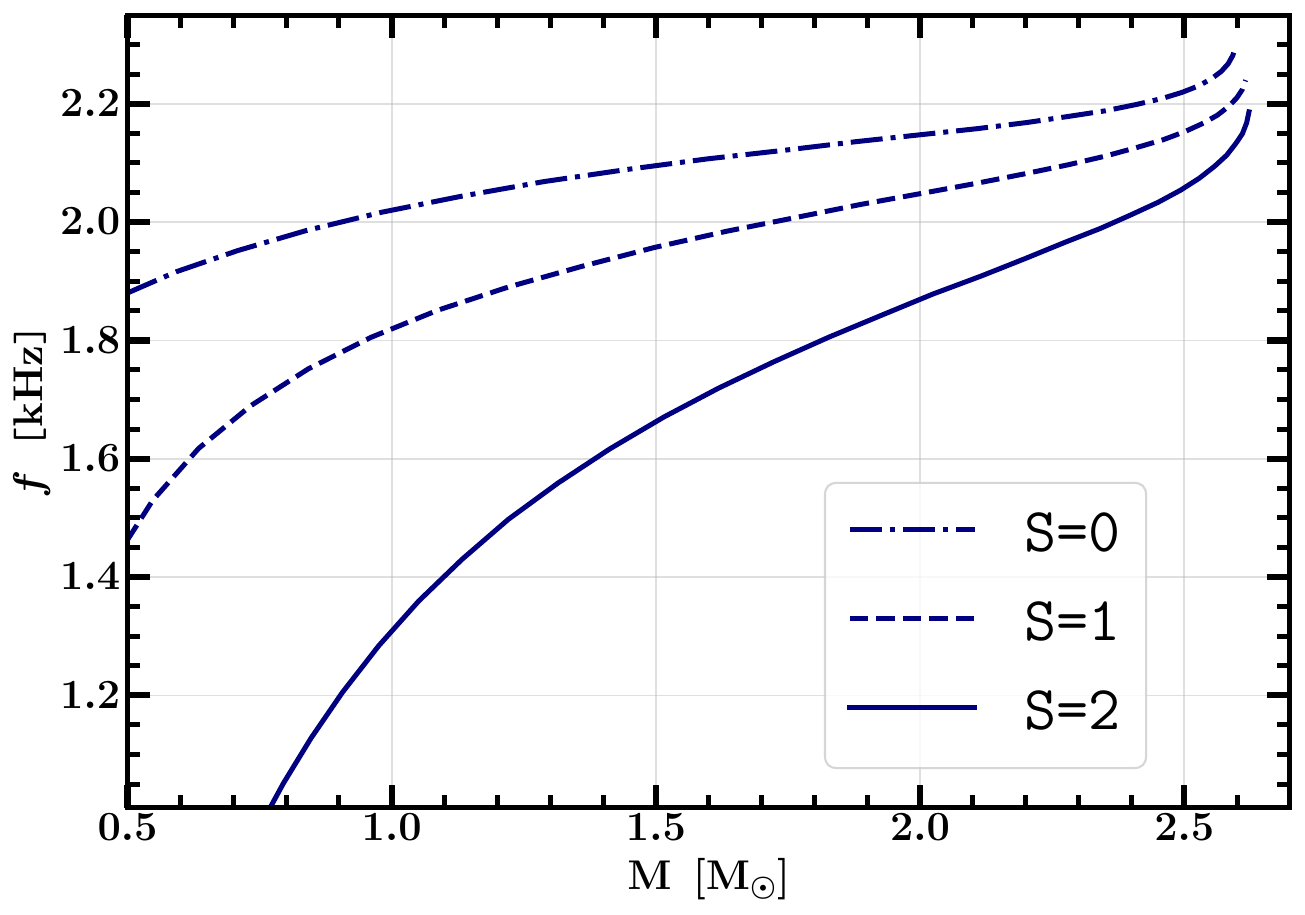}
    \caption{$f$-mode frequencies as a function of NS mass for constant entropy values S=0,1 and 2 ($Y_l$=0.0) shown by respective lines.}
    \label{fmode_s}
\end{figure}

\begin{table*}
\caption{Maximum mass (M), radius (R), dimensionless tidal deformability ($\Lambda$), and $f$-mode frequency ($f$) of the PNS for BigApple parameter set calculated using constant entropy EoS for S = 0, 1 and 2. T$_{Ce}$ denotes the central temperature of the PNS. $\Lambda_{1.4}$ and $f_{1.4}$ are the $\Lambda$ and $f$ values at 1.4M$_\odot$ respectively.}
\centering
\setlength{\tabcolsep}{18.5pt}
\begin{tabular}{ccccccccc}
\hline\noalign{\smallskip} \hline
{\begin{tabular}[c]{@{}l@{}}Entropy\\  (S)\end{tabular}}&
{\begin{tabular}[c]{@{}l@{}}T$_{Ce}$\\
(MeV)\end{tabular}}& 
{\begin{tabular}[c]{@{}l@{}}M\\(M$_\odot$)\end{tabular}}&
{\begin{tabular}[c]{@{}l@{}}R\\(km)\end{tabular}}&
{\begin{tabular}[c]{@{}l@{}}$\Lambda$\\\end{tabular}}&
{\begin{tabular}[c]{@{}l@{}}$\Lambda_{1.4}$\end{tabular}}&
{\begin{tabular}[c]{@{}l@{}}$f$\\
(kHz)\end{tabular}}&
{\begin{tabular}[c]{@{}l@{}}$f_{1.4}$\\
(kHz)\end{tabular}}
\\
\noalign{\smallskip}\hline\noalign{\smallskip}
0&00.000&2.596& 12.128&4.426&642.223&2.309&2.083 \\ 
 1&39.075& 2.615& 12.760&5.389&947.479&2.256&1.934\\
 2& 70.988& 2.622 & 13.464&5.972&1655.316&2.204&1.609 \\  
\noalign{\smallskip}\hline
\hline
\end{tabular}
\label{table:MR_s}
\end{table*}
Fig.\ref{T_mu_p} presents the relationship between the chemical potential ($\mu_n$) and the temperature of a PNS for a fixed entropy. In statistical mechanics and thermodynamics, the chemical potential is a fundamental quantity representing the energy required to add an additional particle to a system. Specifically, for a NS, $\mu_n$ refers to the energy needed to add a neutron to the PNS. From the figure, it is evident that the chemical potential of the system remains nearly constant initially, up to temperatures of 10 MeV (with fixed entropy S=1) and up to 15 MeV (for S=2). However, as the temperature increases, the $\mu_n$ rises rapidly. Notably, the rate of increase in chemical potential is influenced by the lepton fraction, with higher fractions causing a more rapid rise, and lower fractions requiring slightly higher temperatures for the same effect. This sudden increase in chemical potential carries significant implications for the properties and evolution of the NS. It indicates that with rising temperature, more energy is required to add a neutron to the PNS, reflecting the complex interplay of nuclear processes and thermodynamics within the dense environment of the NS.
In Fig. \ref{TD_s012}, we illustrate the correlation between the Tidal deformability ($\Lambda$) and the maximum mass (M) of NSs while considering different entropy values (S = 0, 1, and 2). The figure clearly exhibits that variations in entropy do not significantly alter the overall decreasing trend of Tidal deformability with increasing mass. Nevertheless, entropy does influence the actual values of $\Lambda$. As the entropy increases, so does the Tidal deformability. Notably, in the case where S = 0, the $\Lambda$ values (as presented in Table \ref{table:MR_s}) satisfactorily adhere to the constraints set by two pivotal gravitational wave observational events, namely GW170817 and GW190814.  On the other hand, for the scenarios with S = 1 and S = 2, the $\Lambda$ values only marginally manage to meet these observational constraints. This suggests that NSs with higher entropy values might exhibit different physical properties or internal structures, leading to distinct $\Lambda$ values.
The $f$-mode frequencies of NS play a significant role in understanding their oscillation behaviour and offer valuable insights into the properties of their interior. 
\\
In Fig. \ref{fmode_s}, we investigate these frequencies while keeping the entropy constant.  From the figure, a distinct pattern emerges: as the entropy of the NS increases, the $f$-mode frequency at the NS maximum mass undergoes a decrease. Notably, this trend is particularly evident in the case of lower mass stars with masses below $1.4M_\odot$, where the $f$-mode frequencies experience a significant change of approximately 0.5 kHz as the entropy varies from S=0 to S=1 and so on. The variations in the $f$-mode frequency can be attributed to the alterations in pressure and temperature gradients within the star as the entropy rises. These changes in thermodynamic conditions influence the oscillation modes, resulting in shifts in the $f$-mode frequencies. However, it is important to highlight that at the maximum mass of the NS, the difference in $f$-mode frequencies becomes quite small. This phenomenon arises due to the considerable increase in central density at the maximum mass, leading to the dominance of degeneracy pressure over the effects of entropy in determining the $f$-mode frequency.
\begin{figure}
    \centering
    \includegraphics[height=7cm,width=8.68cm]{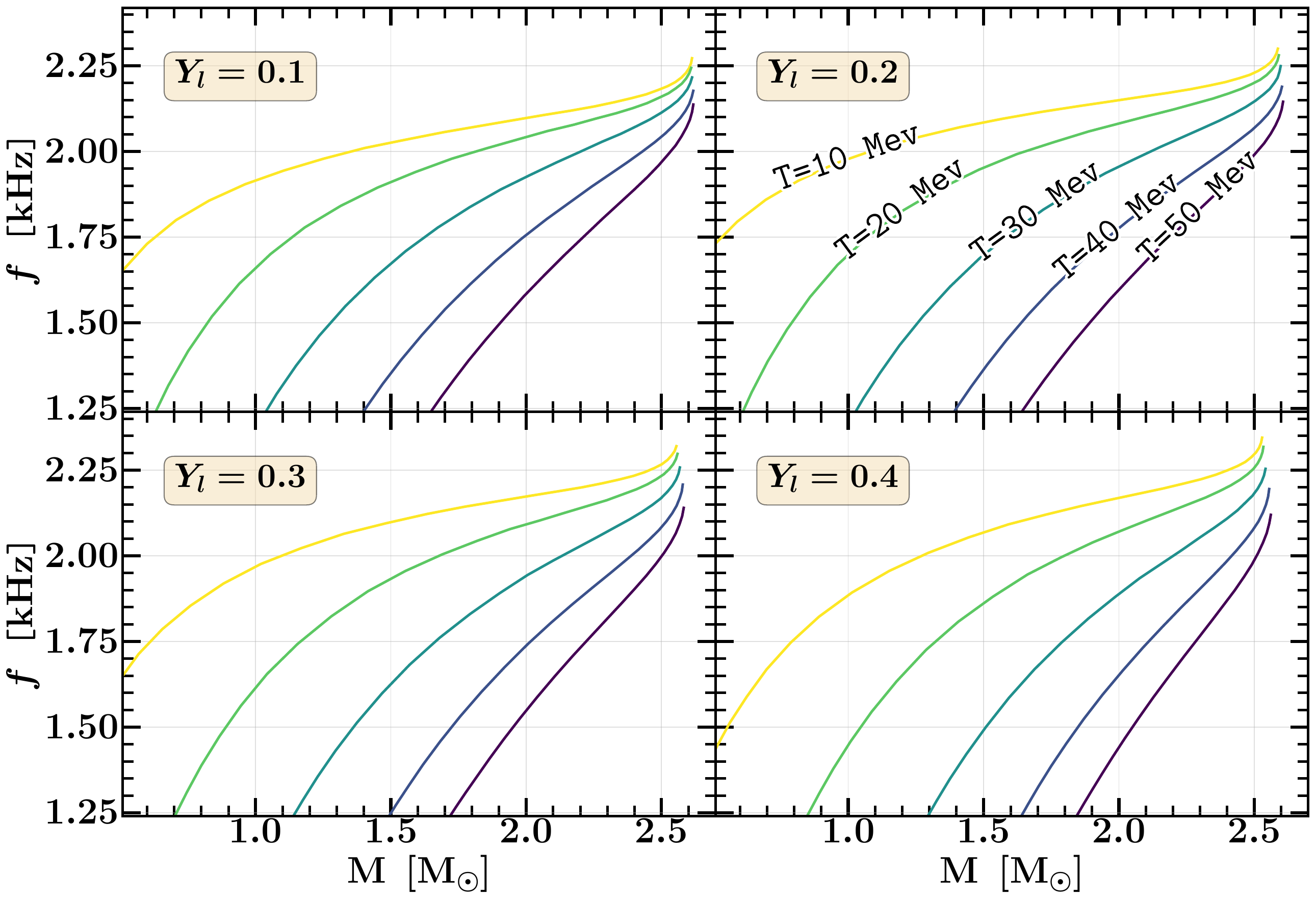}
    \caption{This figure is the same as Fig.\ref{fmode_s}, but with constant temperature EoS at different Temperatures values from 10-50 MeV represented by respective colour lines. The four subplots correspond to values of lepton fraction $Y_l$= 0.1-0.4.}
    \label{fmode_T_all}
\end{figure}
\\
The frequency versus mass plot of the $f$-mode is then examined for the EoS at a constant temperature, while varying the lepton fraction ($Y_l$), see Fig. \ref{fmode_T_all}. Notably, our findings revealed that the $f$-mode frequency corresponding to the maximal mass of the NS decreases as the temperature increases. This finding agrees with prior research~\citep{PhysRevD.107.103054}, which also reported similar trends in $f$-mode frequencies with temperature variations. The change in $f$-mode frequency with increasing temperature can be attributed to the temperature's influence on the structure of the NS. As the star's temperature increases, its thermal energy also increases, leading to a softening of the equation of state. A softer EoS implies that the pressure increases less rapidly with density, making the stellar matter less resistant to compression. As a result, the $f$-mode oscillations encounter a change in their effective potential, resulting in a decrease in their frequency.
The behaviour of $f$-mode frequencies in NSs is influenced by the $Y_l$, temperature, and the EoS and their combined effect. Higher temperature leads to thermal softening of the EoS, which tends to reduce the $f$-mode frequencies. However, the presence of leptons introduces additional pressure support that counteracts thermal softening. Depending on the relative strength of these effects, the $f$-mode frequencies show different behaviours. The interplay between the $Y_l$ and temperature can lead to interesting variations in the $f$-mode frequency and its dependency on the NS's mass. Depending on the specific values of $Y_l$ and temperature, the $f$-mode frequencies exhibit different trends, such as increasing, decreasing, or non-monotonic behaviour with the NS's mass. The observed behaviour of the $f$-mode frequency with increasing $Y_l$ at different temperatures (T) as shown in Fig.\ref{fmode_T_all} can be explained as, at T=10 MeV, i.e. at this low temperature, the thermal energy is minimal, and the EoS is relatively stiff. As the lepton fraction increases, the additional degeneracy pressure from leptons results in a softer EoS, which makes the NS less compact. A less compact star (for a fixed mass, a star will be less compact for a larger radius ($C= \frac{M}{R}$)) allows the $f$-mode oscillations to a larger region of the star, leading to a more extended oscillation period and higher frequency with increasing $Y_l$. At T=50 MeV, i.e. at this higher temperature, the thermal energy becomes more significant, and the EoS is softer. With increasing $Y_l$, the EoS becomes even softer due to the additional contribution from leptons. However, the thermal softening effect becomes more dominant at this temperature. The thermal energy counteracts the degeneracy pressure from leptons, resulting in a decrease in the $f$-mode frequency as $Y_l$ increases.
For other values of temperature, the variation of $f$-mode frequency with lepton fraction is not as clear. At intermediate temperatures, the interplay between the thermal softening and the lepton pressure leads to more complex behaviour. Depending on the specific values of temperature and lepton fraction, the $f$-mode frequencies exhibit different trends or show less clear patterns.

\section{Summary and conclusions}
\label{summary}
In this study, two distinct approaches were employed to investigate the macroscopic properties of PNS using the BigApple parameter set. The first approach maintained a constant temperature across the star, while the second approach held the entropy per baryon at specific values. The constant temperature approach revealed intriguing behavior, particularly in the relationship between entropy per baryon and baryon density. This behavior was attributed to the formation of electron-positron pairs at finite temperatures, which increased the system's entropy even at low baryon densities. The analysis of the mass-radius profile of PNSs showed that including neutrino trapping and temperature considerations had significant effects on the radius and mass of PNSs. The lepton fraction, representing the abundance of leptons like electrons and neutrinos, also played a crucial role in determining the maximum mass and radius of PNSs. Higher lepton fractions resulted in softer equations of state, leading to reduced pressure at a given baryon density and, consequently, smaller maximum masses and larger canonical radii. The study further explored the impact of entropy on PNS properties. Increasing entropy was found to lead to higher maximum masses and flatter mass-radius curves. This effect was attributed to the dominance of pressure support as entropy increased, making the PNS more resistant to compression and thus leading to larger radii for a given mass. The relationship between central temperature and maximum mass of PNSs indicated that more massive stars had hotter cores, influencing various stellar processes such as nuclear reactions and energy generation. Additionally, the mass-radius profile suggested that PNSs tend to contract and shrink in size as they cool down during their evolutionary phases. The study also delved into the chemical potential ($\mu_n$) and its relationship with temperature, revealing that as temperature increased, more energy was required to add a neutron to the PNS. This phenomenon reflected the complex interplay of nuclear processes and thermodynamics within the dense NS environment. Finally, the investigation of the $f$-mode frequencies of NSs showed that these frequencies were influenced by entropy, lepton fraction, temperature, and their combined effects. At the maximum mass of NSs, the difference in $f$-mode frequencies became small due to the dominance of degeneracy pressure over entropy effects. The interplay between these factors led to various trends in $f$-mode frequencies with NS mass and temperature.

\section*{Acknowledgements}
B.K. acknowledges partial support from the Department of Science and Technology, Government of India with grant no. CRG/2021/000101.





\end{document}